\documentclass[english, aps, preprint]{revtex4-1}
\usepackage[T1]{fontenc}
\usepackage[latin9]{inputenc}
\usepackage{graphicx}
\usepackage{amssymb}
\usepackage[english]{babel}
\usepackage{color}
\usepackage{verbatim}
\usepackage[unicode=true, pdfusetitle, bookmarks=true, bookmarksnumbered=false, bookmarksopen=false, breaklinks=false, pdfborder={0 0 1}, backref=false, colorlinks=true]{hyperref}
\hypersetup{linkcolor = blue,  urlcolor  = blue, citecolor = blue, anchorcolor = blue}
\usepackage{breakurl}
\newcommand\micron{{\rm \mu m}}
\newcommand\csurface{\Gamma}
\newcommand\cbulk{c}
\newcommand\thickness{h}

\begin{document}

\title{Surface tension of flowing soap films}
 
\author{Aakash Sane}
\affiliation{School of Engineering, Brown University, Providence, RI 02912, USA}

\author{Shreyas Mandre}
\affiliation{School of Engineering, Brown University, Providence, RI 02912, USA}

\author{Ildoo Kim} 
\email{ildoo_kim@brown.edu}
\affiliation{School of Engineering, Brown University, Providence, RI 02912, USA}

\begin{abstract}

The surface tension of flowing soap films is measured with respect to the film thickness and the concentration of soap solution.
We perform this measurement by measuring the curvature of the nylon wires that bound the soap film channel and use the measured curvature to parametrize the relation between the surface tension and the tension of the wire. 
We find the surface tension of our soap films increases when the film is relatively thin or made of soap solution of low concentration, otherwise it approaches an asymptotic value 30 mN/m.
A simple adsorption model with only two parameters describes our observations reasonably well. With our measurements, we are also able to estimate Gibbs elasticity for our soap film.
 
\end{abstract}

\date{\today}
\maketitle

%%%%
% Introduction
\section{Introduction}
\label{sec:introduction}
% general introduction to soap film channels

Soap film channels have been used as a model system for two-dimensional flow \citep{Kellay1995,Martin:1998ty,Rutgers:1998uz,Vorobieff:1999dd,Jun:2005td,Jung:2006tn,Tran:2010hn,Kim:2015jp} for many years, however, their recent exploitation is expanding in the third dimension.
For example, impingement of solid objects \citep{Courbin:2006vo,Courbin:2006wo,LeGoff:2008dv}, liquid drops \citep{Kim:2010vt,DoQuang:2010tm,Basu:2017wv}, and gaseous jets \citep{Salkin:2016hh} on soap films have been investigated.
The deformation of the soap film accompanies the creation of extra surface area, which sometimes takes a substantial portion of the system's energy.
In this regard, it is critically important to know the surface tension and elasticity of the soap film in order to make the best energetic analysis. 
However our understanding of this property is lacking.

The lack of measurement of the surface tension is due to the technical difficulties that arise from the fact that all existing techniques are not applicable to the soap films.
Some techniques (e.g. pendant drop methods) require the use of a bulk of liquid, whose surface tension is not necessarily equal to that of a soap film, and others (e.g. Du No\"{u}y ring; \cite{duNouy:1925wo}; or Wilhemly plate method) are intrusive and rely on the formation of menisci, which significantly alter the local surface chemistry.
Recent measurement \citep{Adami:2015hy} using a deformable object inserted into soap films is also not suitable if the flow speed of the film is comparable to the Marangoni wave speed \citep{Kim:2017dn}. 

In this work, we present a simple method to measure the surface tension of the flowing soap film.
Our method is applicable to the common setup of a flowing soap film channel \citep{Rutgers:2001wj} without any additional instrumentation except a camera.
We use the fact that those conventional soap films consist of two flexible nylon wires serving as two-dimensional channel walls. 
These flexible wires are tensed by hanging a known weight and curve inwards by the surface tension of the soap film.
The measurement of the surface tension is made possible by measuring the bending curvature and the applied tension. 

We present our measurement of the surface tension of soap films of different thickness and concentration of soap solution.
The thickness of the film is varied from 1 to 10 $\micron$, and three different soap concentrations, 0.5\%, 1\%, and 2\%, are used.
Our measurements show that if the soap film is thick or made from 2\% solution, then it possesses a surface tension of 30 mN/m.
However, if the film is relatively thin or made from 0.5\% or 1\% solutions, its surface tension has a greater value.

Our observation shows the \textit{apparent} equivalence between the thinning and the dilution of the soap film.
The surface tension $\sigma$ depends on the surface concentration $\csurface$ of surfactants, i.e., $\sigma=\sigma_0-\alpha\csurface$, where $\sigma_0$ is the surface tension of pure liquid and $\alpha$ is the proportionality constant \citep{Couder1989,Alberty:1995wb}.
The fact that the thinner films possess higher $\sigma$ implies a smaller value of $\csurface$ than thicker films. 
Dilution of the soap concentration causes the equivalent effect that increases $\sigma$. 
 
We rationalize these observation using a quantitative theory based on the conservation equation for surfactants and the Langmuir's adsorption isotherm.
First, in a patch of a soap film, the total number of surfactants is the sum of surfactants at the interfaces and those in the bulk between interfaces.
This can be written in the form of the conservation law
\begin{equation}
c_0=\cbulk+\frac{2\csurface}{h},
\label{eq:dilution}
\end{equation}
where $c_0$ is the concentration of the soap solution and $\cbulk$ is the concentration of interstitial surfactants.
Second, using the Langmuir's adsorption isotherm, the relation between $\csurface$ and $\cbulk$ is formulated with two parameters $\csurface_\infty$ and $\cbulk^*$
\begin{equation}
\csurface= \frac{\csurface_\infty }{1+\cbulk^*/\cbulk}.
\label{eq:langmuir}
\end{equation}
Physically, $\csurface_\infty$ is the concentration of surfactants when the surface is fully occupied by surfactants, and $\cbulk^*$ is the value of $\cbulk$ when $\csurface=\csurface_\infty/2$.
Using our measurement of $\sigma$ with respect to $c_0$ and $h$, we estimate these two parameters $\csurface_\infty$ and $\cbulk^*$.
Our analysis shows that this two-parameter model describes our system reasonably well.
The two parameter model also enables us to estimate the Gibbs elasticity for our soap film setup.

% Experimental Method
\section{Experimental Method}

% 1. soap film
\subsection{Soap film setup}

\begin{figure}
\begin{centering}
\includegraphics[width=9cm]{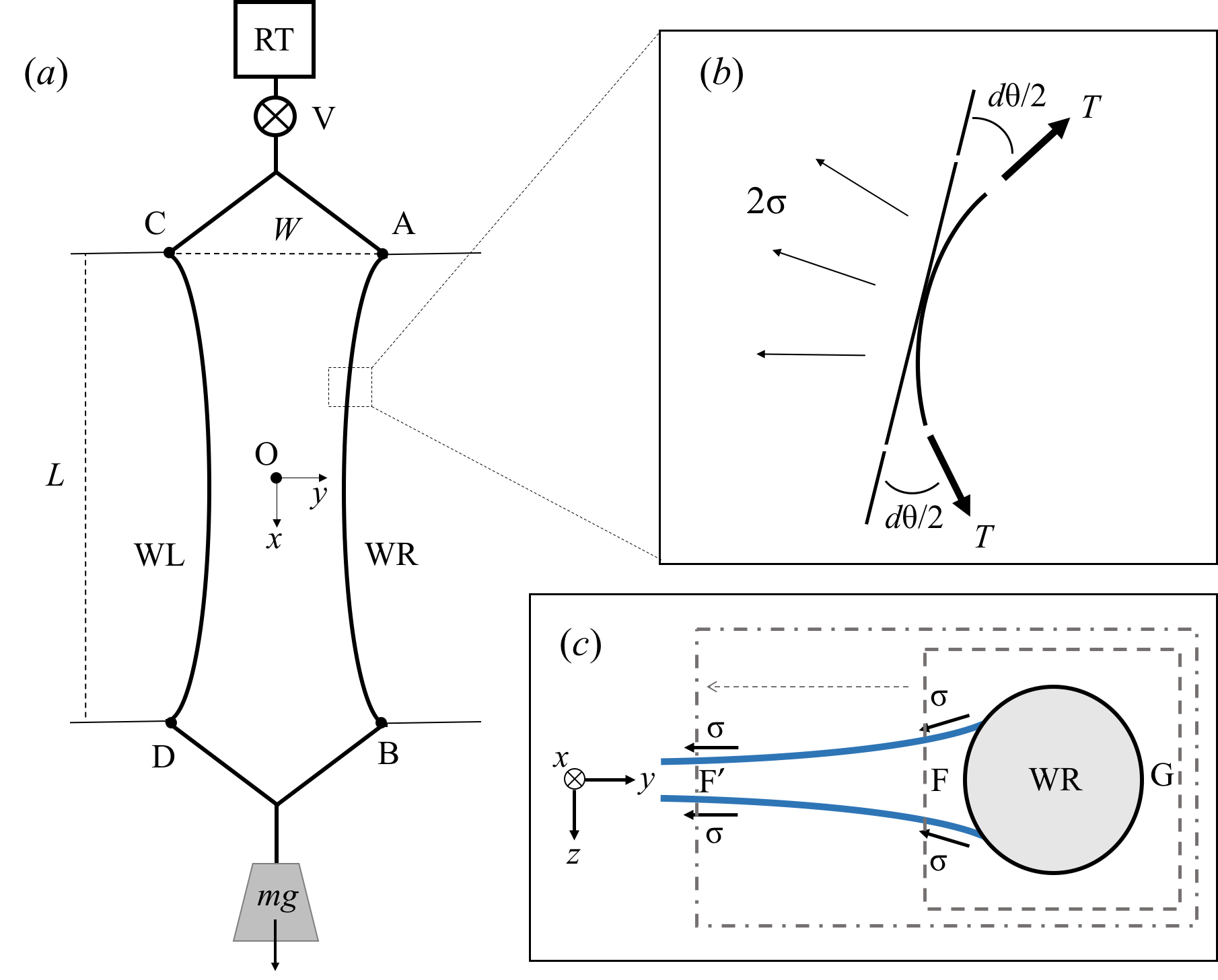}
\par\end{centering}
\caption{
(\textit{a})
The main component of the standard soap film channel setup is two flexible nylon wires (WL and WR) that are anchored at four points, A, B, C, and D.
In the presence of a soap film, wires are bent toward each other due to the surface tension.
(\textit{b})
A length element of wire $ds$ is made tight by $T$ and bent by $\sigma$.
The factor 2 of $\sigma$ reflects that the soap film has two surfaces.
(\textit{c})
Cross section of the film in $y$-$z$ plane. 
In the control volume shown by dashed rectangle, pressure at point F is different from the pressure at point G due to the curvature in $z$ direction.
However, this pressure difference is canceled out with the effect of the menisci.
This can be shown by extending the control volume to point $\rm F'$ where the film is flat.
}
\label{fig:apparatus}
\end{figure}

We carry out the experiments using a standard soap film setup, which is introduced and discussed in published work \citep{Rutgers:2001wj, Kim:2017dn}.
The main component of the standard setup is a pair of nylon wires (WL and WR in Fig. \ref{fig:apparatus}(a)) that are connected to a soap solution reservoir at the top (RT) and to a suspended weight $mg$ at the bottom. 
The suspended weight creates tension in the wires which causes them to remain vertical. 
To create a soap film, we open a needle valve (V) to allow soapy water to flow along two nylon wires as they are initially abreast, and then we pull them apart from each other by using four auxiliary wires.
The auxiliary wires provide support so that the main wires are fixed at four points: A, B, C, and D.
We set the coordinate system such that $\hat{x}$ is longitudinal and $\hat{y}$ is transverse to the flow with the origin O being the centre of the soap film. The soap film channel is carefully tuned to be symmetric about both the axes. 

%%% Figure 1

We prepare solutions of commercial dish soap (Dawn, P\&G) in distilled water at three different concentrations.
The soap concentration $c_0$ is defined as the volume fraction of the soap, with $c_0=$0.5\%, 1.0\%, and 2.0\% being used for the current study.
The flow rate $\phi$ of the channel is varied from 0.26 to 1.55 $\rm cm^3/s$ by adjusting the needle valve (V).

The whole channel is 1.8 m long, and the length $L$ between two fixed points A and B is 1.2 m and is equal to the distance between C and D, i.e., $L\equiv\overline{{\rm AB}}=\overline{{\rm CD}}=1.2 \, \rm m$.
Likewise, the reference width of the channel $W$ is defined as $W\equiv\overline{{\rm AC}}=\overline{\rm BD}$ and varied from 6 cm to 14 cm.
We use a mass of 400 g for the hanging weight at the bottom, therefore a tension of 1.96 N is equally applied to WL and WR.
The even distribution of the weight is confirmed by observing that the wires oscillate at the same frequency when plucked.
In this setting, the applied tension $T$ is $\mathcal{O}(10^0)$ N, the surface tension $\sigma$ is $\mathcal{O}(10^{-2})$ N/m, and the length of the soap film $L$ is $\mathcal{O}(1)$ m.
The channel width is then contracted slightly, by 1\% of its original width.
We tune the weight to be small enough to substantiate the width contraction but yet to minimize the distortion of the channel that may disturb the flow.

\begin{figure}
\begin{centering}
\includegraphics[width=\columnwidth]{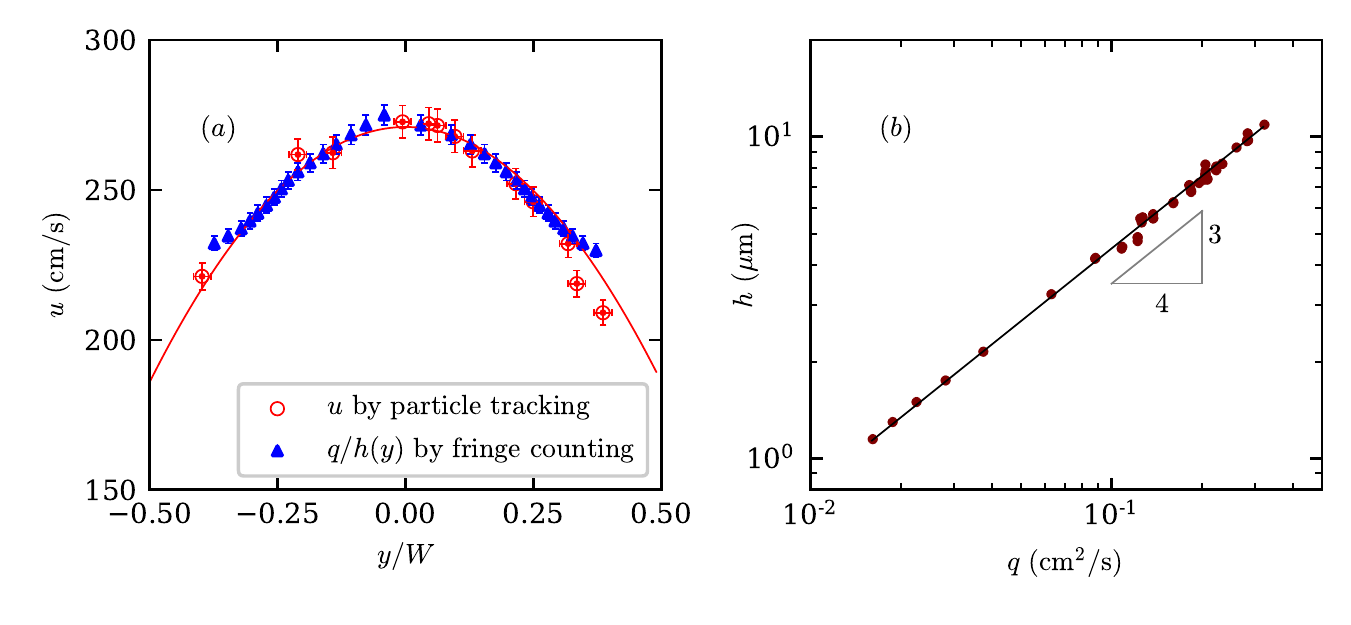}
\par\end{centering}
\caption{
(\textit{a}) The flow speed profile across $y$ at $q=0.33\,\rm cm^2/s$. 
The open circles are the direct measurement of $u(y)$ using the particle tracking, and the closed squares are the calculation of $q/h(y)$, where $q$ is fixed and $h(y)$ is measured by counting fringes of equal thicknesses.
(\textit{b}) The relation between $\thickness$ and $q$. 
We find that $\thickness= h_0(q/q_0)^{0.75}$, with $q_0 = 0.1\,\rm cm^2/s$ and $h_0 = 4.5 \,\micron$.
}
\label{fig:flowprofile}
\end{figure}

In developing a quantitative theory, it is necessary to find the thickness of the soap film.
We define the \textit{mean} thickness of the film as $\thickness\equiv q/u_t$, where $q\equiv\phi/W$ is the flux per unit width and $u_t$ is the terminal velocity of the flow.
We find that $q$ is independent of $y$.
Using the particle tracking velocimetry, we find that the profile of $u(y)$ is parabolic as seen in Fig. \ref{fig:flowprofile}(a).
This profile matches well with an independent measurement $q/h(y)$ counting fringes of equal thicknesses under a monochromatic illumination. 
Also, $q$ only weakly depends on $x$.
Even though the channel width depends on $x$, its variation is limited to a few percent and smaller than 5\% uncertainty of $\phi$.

%%% Figure 2

We observe that both $\thickness$ and $u_t$ are increased by increasing $q$, namely $\thickness\propto q^{0.75} $ and $u_t\propto q^{0.25}$.
Here, the terminal velocity $u_t$ is measured at $y=0$ and approximately 0.7 m away from the top of the channel where it becomes independent of $x$. 
This gives a relation of $\thickness$ to $q$ as delineated in Fig. \ref{fig:flowprofile}(b),
\begin{equation}
\thickness= \thickness_{0} \left(q/{q_{0}}\right)^{0.75},
\label{eq:film_thickness}
\end{equation}
where $q_0=0.1\,\rm cm^2/s$ and $\thickness_0=4.5 \,\rm \mu m$.
The power exponent 0.75 in Eq. (\ref{eq:film_thickness}) is somewhat higher than 0.6 reported in literature \citep{Rutgers:1996vg}.
Currently, no theory is available to take account of this scaling relation. 
However, we find that our empirical relation between $q$ and $\thickness$ is clear and reproducible; in the rest of the paper, we use Eq. (\ref{eq:film_thickness}) to estimate $\thickness$. 
In our range of $\phi$ and $W$, $q$ is varied from 0.02 to 0.4 $\rm cm^2/s$, and $h$ is varied from 1 to 12 $\micron$.

% 2. relations
\subsection{Relation of the surface tension to the curvature}

When a soap film is formed, the surface tension of the fluid pulls the wires (WL and WR) toward each other and creates curvature.
In the following calculation, we consider the force balance on a length element $ds$ of the wire as seen in Fig. \ref{fig:apparatus}(b).

In the tangential direction, the forces acting on $ds$ are the following: the tension $T$ of the wire, the gravitational force, and the viscous force between the wire and the flow.
Here we assume that the deflection of the wire is so small that the wire is nearly vertical. 
This assumption allows us to approximate the gravitational force to be in the tangential direction.
When the wire is stationary,
\begin{equation}
\frac{\partial{T}}{\partial{x}} + \rho \pi \frac{d^2}{4} g -\mu \frac{\partial{u}}{\partial{y}} \pi d =0,
\end{equation}
where the diameter of the wire $d=0.038 \,\rm cm$, the density of the wire $\rho=1.15 \,\rm g/cm^3$, and the viscosity of the soap solution $\mu\simeq8.9\times 10^{-3}\,\rm Pa\cdot s$.
Considering that $du/dy\approx 10^{3} \,\rm s^{-1}$ is usually reported in soap films \citep{Rutgers:1996vg}, we find that $\partial T / \partial x \approx 10^{-4}$ \,\rm N/m.
We conclude that $T=mg/2=1.96 \,\rm N$ at any location of the wire.

In the normal direction, the surface tension $\sigma$ exerts to bend the wires, and the tension $T$ resists as seen in Fig. \ref{fig:apparatus}(b).
Using the Euler-Bernoulli equation,
\begin{equation}
E I \frac{\partial^4 y}{\partial x^4} - 2\sigma + T \kappa=0,
\label{eq:normal_force_balance_1}
\end{equation}
where $E\approx 10^9 \,\rm Pa$ is the elastic modulus of the wire, $I=\pi (d/2)^4 /2 \simeq 2\times 10^{-15} \,\rm m^4$ is the second moment of area, and $\kappa=d\theta/ds\approx d^2y/dx^2\simeq (30 \,{\rm m})^{-1}$ is the curvature.
Comparing the relative importance of the first term to the second term as $EI\kappa/(2\sigma L^2) \approx 10^{-6}$ and to the third term as $EI/(TL^2) \approx 10^{-6}$, we conclude that the bending stiffness of the wire is negligible.
Eq. (\ref{eq:normal_force_balance_1}) is then simplified to
\begin{equation}
\sigma = \frac{T\kappa }{2} = \frac{mg}{4} \frac{\partial^2 y}{\partial x^2},
\label{eq:tensionbalance}
\end{equation}
which provides us the means to measure $\sigma$ by characterizing the local curvature of the wire.

We also examine the effect of the Laplace pressure due to the curvature in $z$ direction.
Figure \ref{fig:apparatus}(c) shows the cross section of the film in $y$-$z$ plane.
As shown in Fig. \ref{fig:flowprofile}(a), the film is thicker near the wires, and there is non-zero curvature in $z$ direction.
In a control volume enclosed by dashed rectangle in Fig. \ref{fig:apparatus}(c), the pressure at point F is smaller than the pressure at G, and exerts an additional force to bend the wire. 
However, the curvature also reduces the component of surface tension and increases the area of the control surface.
Because the Laplace pressure is related to the curvature, these effects cancel each other exactly.
This becomes clear when we extend the control volume to point $\rm F'$ (shown in dash-dotted line) where the film is flat, and the net force to the control volume is zero except $\sigma$.
Combined with the observation that there is no flow in the $y$ direction in the soap film, it is inferred that Eq. (\ref{eq:tensionbalance}) holds irrespective of the control volume chosen.

% 3. measurement
\subsection{Measurements}

A digital camera is set for visual observation of the soap film. 
A conventional digital single lens reflection camera (Nikon D90) is placed approximately 3.5 m away from the soap film.
The camera's optical axis is aligned with $z$ axis of our coordinate system so that the centre of the image is the centre of the soap film channel (the origin in the coordinate system). 
The entire soap film is captured in a single frame.

The captured digital image of the soap film channel is processed using our in-house code for the digitization and regression analysis.
Each photo is converted to a set of $(x, y)$ coordinates of points on the wire, and more than 3000 data points are acquired per wire.

We then fit the digitized data to the polynomial function $y=\sum a_n x^n$.
Inspection of multiple cases shows that $a_n$ for $n>2$ is negligible, and we conclude that the use of the quadratic regression model $y=a_2 x^2+ a_1 x+ a_0$ is sufficient to fit our data. 
The coefficients $a_2$, $a_1$, and $a_0$ are determined by least-square method, and the curvature $\kappa$ is calculated by taking a second derivative, as $\kappa={\partial^2 y}/{\partial x^2}=2a_2$.
Then, $\sigma=mga_2/2$ according to Eq. (\ref{eq:tensionbalance}).

% Result and Discussion
\section{Results and Discussion}

\subsection{Surface tension}

\begin{figure}
\begin{centering}
\includegraphics[width=\columnwidth]{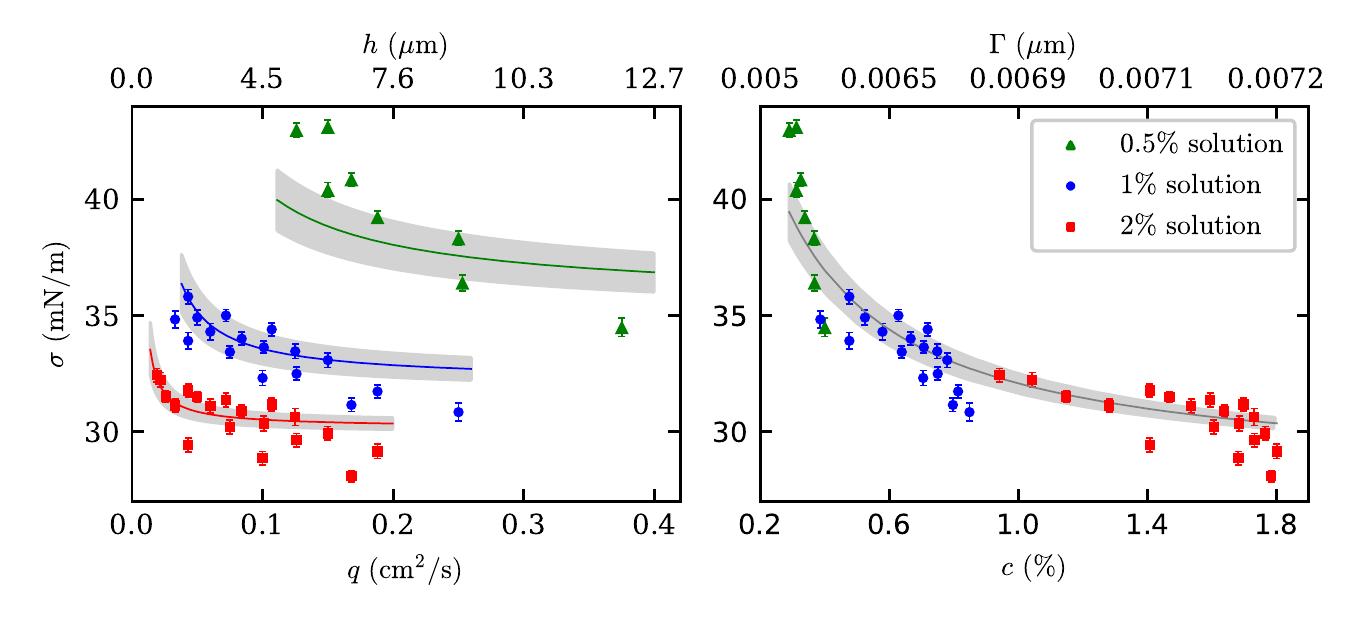}
\par\end{centering}
\caption{
(\textit{a})
The surface tension $\sigma$ measured using Eq. (\ref{eq:tensionbalance}).
When $c_0$ is high, $\sigma$ is independent of $h$. 
However, when $c_0$ is low, $\sigma$ increases as $h$ increases.
(\textit{b})
Data points from various experimental conditions all collapse into a single curve that represents the equilibrium between $\csurface$ and $\cbulk$.
It is inferred that the main surfactant of our soap films are in equilibrium.
}
\label{fig:surfacetension400g}
\end{figure}

In Fig. \ref{fig:surfacetension400g}(a), our measurements of $\sigma$ are presented with respect to $q$, ranging from 0.02 to 0.4 $\rm cm^2/s$, for three different $c_0=$ 0.5\%, 1\%, and 2\%.
When $q$ is sufficiently large thus the film is sufficiently thick, $\sigma$ approaches an asymptotic value $\sigma_\infty= 30\,\rm mN/m$.
This value of $\sigma _\infty$ equals to the surface tension of bulk solution; in a separate measurement of $\sigma$ of soap solutions in Petri dishes using du Nu\"{o}y ring method, we find that $\sigma=30\,\rm mN/m$ if $c_0>0.05\%$.
This observation implies that in the asymptotic regime, the surface of the soap film is fully covered with surfactants and the surface concentration $\csurface$ approaches a constant value $\csurface_\infty$ as $\sigma$ also approaches $\sigma_\infty$.
Using the asymptotic properties, the proportionality constant in the assumed linear relation $\sigma=\sigma_0 -\alpha \csurface$ can be obtained as $\alpha= (\sigma_0 - \sigma_\infty)/\csurface_\infty$.

The required thickness to possess $\sigma_\infty$ is inversely proportional to $c_0$.
We find that $\thickness=3.8\,\rm \mu m$ ($q=0.08 \,\rm cm^2/s$) is required for $c_0=2\%$, and $\thickness=7.6\,\rm\mu m$ ($q=0.2\,\rm cm^2/s$) is required for $c_0=1\%$ to possess $\sigma_\infty$. 
By extrapolating data points, $\thickness\simeq 15 \,\rm\mu m$ ($q=0.5\,\rm cm^2/s$) is estimated to be required for $c_0=0.5\%$.

%%% Figure 3

The inverse proportionality between $c_0$ and $h$ shows the apparent equivalence between thinning and dilution.
Our measurement in Fig. \ref{fig:surfacetension400g}(a) indicates that there are two ways to decrease $\csurface$ in soap films; one could decrease $\csurface$ by reducing $c_0$ (dilution), or by reducing $h$ (thinning).
If we consider a patch of a soap film, thinning does not change the volume of the patch but introduces fresh interface and reduces the number density of surfactants in the patch. 
The dilution results in the reduction of the total number of surfactants in the patch and $\csurface$ is reduced.

To accommodate our observation quantitatively, we present a self-consistent model of the surface tension by using the conservation equation (or as we call it \textit{the surface dilution equation}) in Eq. (\ref{eq:dilution}) and the Langmuir adsorption equation in Eq. (\ref{eq:langmuir}). 
The basic feature of the Langmuir adsorption equation is that $\csurface$ increases with $c$ for low concentration, and then reaches an asymptotic value $\csurface_\infty$. 
This behaviour qualitatively agrees with that of soap solutions; when $c_0$ is small, $\csurface$ and $c$ are roughly proportional to each other, for asymptotically large $c_0$, $\csurface$ reaches a constant asymptotic value \citep{Tajima:1970vl}.
Leaving two parameters $\csurface_\infty$ and $c^*$ to be determined later, we solve for $\csurface$ using Eqs. (\ref{eq:dilution}) and (\ref{eq:langmuir}),
\begin{equation}
\csurface=\csurface_\infty - \csurface'(h,c_0; \csurface_\infty, c_b^*),
\label{eq:cs}
\end{equation}
where $\csurface'(h,c_0; \csurface_\infty, c_b^*)\equiv {[(c_0h+c^*h - 2\csurface_\infty)^2/16 + \csurface_\infty c^* h/2 ]}^{1/2}-(c_0h+c^*h-2\csurface_\infty)/4$.
The function $\csurface'$ is positive-valued and approaches zero as $h\rightarrow\infty$.
Then the surface tension $\sigma$ is
\begin{eqnarray}
\sigma&=&\sigma_0-\alpha\csurface=\sigma_\infty + (\sigma_0-\sigma_\infty) \frac{\csurface'(h,c_0)}{\csurface_\infty},
\label{eq:surfacetensionmodel}
\end{eqnarray}
where $\sigma_\infty\equiv\sigma_0-\alpha\csurface_\infty$ equals the experimentally observed value 30 mN/m.

We now determine two parameters $\csurface_\infty$ and $c^*$ by regression.
Experimental data in Fig. \ref{fig:surfacetension400g}(a) are provided as an input to the model in Eq. (\ref{eq:surfacetensionmodel}), and the model is iteratively solved to minimize the mean squared residual of the fitting.
The analysis yields that $\csurface_\infty=0.76 \%\cdot\micron$ and $\cbulk^*=0.10\%$ as the best fitting of our data.
In Fig. \ref{fig:surfacetension400g}(a), the calculation of the model using the obtained parameters are displayed as solid lines.
The grey shades around the curve shows $\pm$10\% uncertainty of the calculation.
By substituting $\csurface_\infty=0.76 \%\cdot\micron$ and $\cbulk^*=0.10\%$ in Eqs. (\ref{eq:cs}) and (\ref{eq:dilution}), $\csurface$ and $\cbulk$ are calculated for each measured datum.
Fig. \ref{fig:surfacetension400g}(b) shows $\sigma$ plotted with respect to such calculated $\cbulk$, and our measurement from three different concentrations collapses into a single curve.
The black line in the figure is the Langmuir adsorption in Eq. (\ref{eq:langmuir}) and represents the relation between $\csurface$ and $\cbulk$ in equilibrium. 

Two important implications about our soap films are learned.
First, the primary surface-active component of our soap films approach equilibrium.
The commercial dish soap we use is a mixture of several species of surfactants with different properties. 
However, it is shown in Fig. \ref{fig:surfacetension400g}(b) that the data points from various experimental conditions all collapse into the equilibrium relation, and this collapse strongly indicates that $\csurface$ and $\cbulk$ of the main surfactants are in equilibrium in our soap films. 
Further, we estimate the diffusive time scale across the thickness $\tau_d\simeq \thickness^2/6D\approx $ 0.4 to 40 ms using the diffusion coefficient $D\simeq 4\times 10^{-6} \,\rm cm^2/s$ \citep{Couder1989}.
The flow speed $u_t\sim$ 2.5 m/s, and the longitudinal distance to reach equilibrium, $u_t \tau_d$, is less than 0.1 m. 
Considering that our soap film has an initial expansion zone that is approximately 0.3 m long, it is evident that the equilibrium between the surface and the bulk must have been reached even in the thickest film.
Second, however, other than the primary surfactants, there may be other, perhaps heavier, components of the soap that remain out of equilibrium in the soap film setup. 
This is suuported by our independent surface tension measurements of soap solutions, which show further reduction in the value of the surface tension than that predicted by the Langmuir isotherm.

\subsection{Gibbs elasticity}

Elasticity is related to $\sigma$ and is pertinent for the energetic analysis of soap films. 
The elasticity of a soap film quantifies the stability of soap film under disturbance and is defined as the change in $\sigma$ per the fractional change in the surface area, i.e., $E=Ad\sigma/dA$, where $A$ is the area of a soap film \citep{Lucassen:1970ta}.
The measurement of $\sigma$ with respect to $\thickness$ allows us to calculate the Gibbs elasticity of soap films.
Using the incompressibility $\ln A =-\ln \thickness$, the elasticity is
\begin{equation}
E=-\frac{d\sigma}{d\ln\thickness}.
\label{eq:gibbs_elasticity}
\end{equation}
Depending on the time scale of the disturbance, there are two elasticities of soap films, the Marangoni and the Gibbs elasticity.
The Marangoni elasticity applies to the scenario in which a soap film is stretched suddenly, i.e. the time scale of the disturbance $\tau$ is shorter than the diffusive time scale $\tau_d$, so that the interstitial surfactants do not have time to diffuse to the surfaces. 
Then the local decrease in $\csurface$ will increase $\sigma$, producing Marangoni stress that recovers the soap film back to the previous equilibrium state.
Otherwise, $\tau>\tau_d$, the Gibbs elasticity emerges when a patch of a soap film is stretched slowly so that the interstitial surfactants diffuse to the surfaces and a new thermodynamic equilibrium between $\csurface$ and $\cbulk$ is reached.
In current study, we measure $\sigma$ with respect to $\thickness$ while $\csurface$ and $\cbulk$ of the main surfactants remain in equilibrium, and therefore the Gibbs elasticity can be measured.

\begin{figure}
\begin{centering}
\includegraphics[width=10cm]{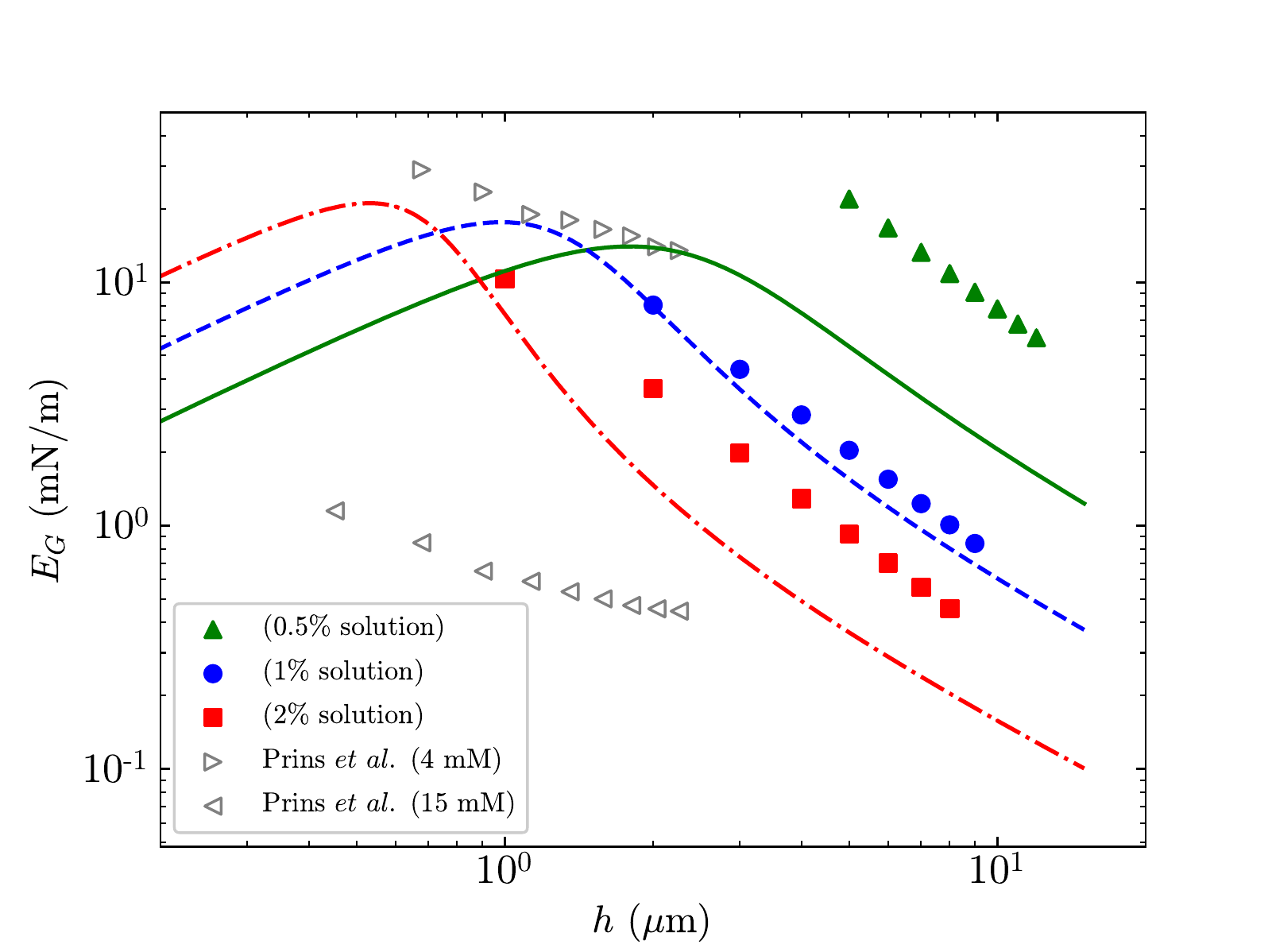}
\par\end{centering}
\caption{
Gibbs elasticity is measured by extracting $ - d \sigma/d \ln h$ from Fig. \ref{fig:surfacetension400g} with Eq. (\ref{eq:gibbs_elasticity_expand}).
The measured values are consistent with the reported values in literatures \citep{Prins:1967tk} but greater than the prediction of Langmuir adsorption model in Eq. (\ref{eq:langmuir}).
}
\label{fig:Gibbs_elasticity}
\end{figure}

Our firsthand inspection of $\sigma$ with respect to $\ln\thickness$ suggests that $E_G$ is proportional to $\thickness^{-3/2}$ for a given $c_0$.
From the observation, we make an ansatz that helps us to pick $E_G$ from the noisy data such that 
\begin{equation}
E_G(c_0,\thickness)\simeq  E_0(c_0) \thickness^{-3/2},
\label{eq:gibbs_elasticity_expand}
\end{equation}
where $E_0$ is the concentration dependent parameter.
Integration of Eq. (\ref{eq:gibbs_elasticity}) using Eq. (\ref{eq:gibbs_elasticity_expand}) gives that 
\begin{equation}
\sigma=\sigma_\infty+ \frac{2}{3} E_0 \thickness^{-3/2}.
\label{eq:gibbs_elasticity_integrated}
\end{equation}
Measurement of $E_G$ is possible by fitting the data using Eq. (\ref{eq:gibbs_elasticity_integrated}).
We get $E_0=10.3$ for soap films made of 2\% soap solution, $E_0=22.8$ for 1\% soap films, and $E_0=246$ [mN/m$\cdot{\mu\textrm{m}}^{3/2}$] for 0.5\%.
All cases, $\sigma_\infty=30\pm1\,\rm mN/m$ from the fitting.
In the range of $h$ where Eq. (\ref{eq:gibbs_elasticity_expand}) is valid, our measurement of $E_G$ is displayed in Fig. \ref{fig:Gibbs_elasticity}.

%%%% figure 4

In Fig. \ref{fig:Gibbs_elasticity}, we compare our measurement with the values in the literature.
Measurement by \citet{Prins:1967tk} reports that $E_G$ is 14 to 29 mN/m for 4 mM, and 1.1 to 0.45 mN/m for 15 mM solution of sodium dodecyl sulfate (SDS)\footnote{In Ref. \citet{Prins:1967tk}, the \textit{film} elasticity, which is defined as $E_f=2Ad\sigma/dA$ is measured. We used the conversion that $E_G=E_f/2$.}.
According to the manufacturer's chemical safety data sheet, the main ingredient of the soap that we used is also SDS, and the concentration of SDS in 0.5\% solution is estimated roughly $5\pm2$ mM. 
The comparison shows that our measurement of $E_G$ is in rough agreement with the previous reports and supports our measurements.

The curves in Fig. \ref{fig:Gibbs_elasticity} are the estimation of $E_G$ based on the Langmuir adsorption through direct differentiation of Eq. (\ref{eq:surfacetensionmodel}).
For all $c_0$, our physical model underestimate the elasticity.
The discrepancy between the measurement and the model shows the complex fluid nature of the soap films.
The Langmuir adsorption is a simple and idealized model in which a constant adsorption and desorption rate is assumed.
However, in soap films, the interplay between the surface chemistry and the hydrodynamics complicates the overall dynamics, and a complete description of its physical property requires further thorough consideration.

Finally, we remark that in the range that we can conveniently establish a flowing soap film, the Gibbs elasticity is smaller than the Marangoni elasticity.
The Marangoni elasticity is reported to be 22 mN/m in soap films made of 1 to 4\% soap solution in the thickness range between 4 and 11 $\rm \mu m $ \citep{Kim:2017dn}.
Even though a soap film requires both elasticities in order to last for a long time, only those with sufficient Marangoni elasticity can be established at the first place. 
Also, two elasticities are complementary; the Gibbs elasticity diminishes as $\csurface\rightarrow\csurface_\infty$, the Marangoni elasticity diminishes as $\csurface\rightarrow0$. 
Considering that the fast disturbance is commonly encountered in a usual lab environment, soap films with greater Marangoni elasticity is expected to be observed more frequently.

% Summary
\section{Summary}

We presented a non-intrusive method to measure the surface tension of a flowing soap film setup.
The method is applicable to standard soap film setups whose main component is thin flexible wires serving as channel walls.
When a soap film is formed between wires, the wires are bent toward each other by the action of the surface tension. 
We showed that the bending curvature is determined by the relative strength of the surface tension to the tension of the wire. 

Using the presented protocol, the surface tension was measured by probing the bending curvature of soap films made from different conditions. 
Our measurements show that a soap film has a surface tension of 30 mN/m if its thickness is relatively thick or if it is made of soap solutions of higher concentrations. 
Otherwise, the surface tension deviates from the asymptotic value 30 mN/m and increases.
Two distinct physical processes, thinning of soap film and dilution of soap solution, yield the same consequences that increases the surface tension.

We demonstrated a theoretical model using surfactant conservation and Langmuir adsorption isotherm. 
These two equations can be solved for an analytic solution, and two parameters of the model determined by matching the model and experimental data. 
These parameters are in agreement with other independent measurements.

Lastly, using our measurement of the surface tension with respect to the thickness of the film, we estimated the Gibbs elasticity.
In our experimental range, the Gibbs elasticity is greater when the film is made of dilute soap solution or when the film is thinner.
In our setup, the Marangoni elasticity is bigger than the Gibbs elasticity.

% Acknowledgement
\section*{Acknowledgement}

S. M. and I. K. contributed equally.

% Bibliographical information
\bibliographystyle{apsrev4-1}
\bibliography{soapfilm2017}

\end{document}